# CONCEPTS AND EVOLUTION OF RESEARCH IN THE FIELD OF WIRELESS SENSOR NETWORKS

Ado Adamou ABBA ARI [1, 3 *], Abdelhak GUEROUI [1], Nabila LABRAOUI [2] and Blaise Omer YENKE [3]

[1] PRISM, University of Versailles St-Quentin-en-Yvelines, France
[2] STIC, University of Tlemcen, Algeria
[3] LASE, University of Ngaoundere, Cameroon

*ABSTRACT*

*The field of Wireless Sensor Networks (WSNs) is experiencing a resurgence of interest and a continuous evolution in the scientific and industrial community. The use of this particular type of ad hoc network is becoming increasingly important in many contexts, regardless of geographical position and so, according to a set of possible application. WSNs offer interesting low cost and easily deployable solutions to perform a remote real time monitoring, target tracking and recognition of physical phenomenon. The uses of these sensors organized into a network continue to reveal a set of research questions according to particularities target applications. Despite difficulties introduced by sensor resources constraints, research contributions in this field are growing day by day. In this paper, we present a comprehensive review of most recent literature of WSNs and outline open research issues in this field.*

## 1. INTRODUCTION

During last decade, the field of WSNs has attracted the attention of scientific and industrial community. With this particular kind of ad hoc networks, it is possible to perform various applications grouped into monitoring and tracking of some activities. The rapid evolution of the Micro-Electro-Mechanical Systems (MEMS) has contributed to the development of small and smart sensors [1]. These sensors have become increasingly very small in terms of size, more intelligent and less expensive [2]. A node in WSN consists of a sensor unit, a processing and data storage unit, a wireless transmission module and a power management unit. Each node is able to gather and process physical information in order to transmit these data to a base station or sink node. WSN consists of a deployment of one or more sink nodes and a number of sensor nodes in a physical environment.

Wireless sensors are designed with huge resource constraints: a limited amount of energy; reduced computing capacity; limited memory size and storage; short-range of communication and reduced bandwidth. So, it appears some problems in networks architectures, QoS (Quality of Service), coverage, security, fault tolerance, etc. [3]. In a WSN, energy consumption depends of network architecture, environment in which the network is deployed and the underlying application. WSNs have many applications in environmental monitoring, prevention of natural disasters, military sector, in the medical, bio-medical and veterinary field, in commercial area, especially in supply chains, aviation and automotive safety, in field of distribution of energy and in agriculture [5, 6, 7, 8, 9, 10].

In general, research papers present specific results or reviews of specific research area. The novice who is engaged in the study of WSNs does not have a panoramic view of the ongoing and forthcoming works in the field of sensor networks. It is, therefore, important to provide an overview of main concepts, and also the evolution of the research. The main motivation of this paper is to provide a comprehensive overview of the field of WSNs, its evolution and actual research issues.

The research WSN domain is constantly evolving as evidence by publication of several contributions, but improvement is still possible and some challenges remain open: location, timing, coverage, energy management, security, synchronization aggregation and data compression. This paper sets out to present a brief survey in the field of WSN.

The rest of this paper is organized as follow: in section 2, we provide some most recent survey of WSNs and highlight the originality of this review, Section 3 present sensors and types of sensor networks; in section 4, we discuss on architectures, offered services and fault tolerance; in section 5, some practical applications of sensor-based network are presented; section 6 present a review of some communication protocols and a comparison of them is proposed; section 7 describes sensor network security and some challenges are introduced; section 8, discusses on open research issues and section 9 concludes the paper.

## 2. RELATED WORK

A top-down approach is followed by authors in [61], in order to give an overview of several applications of sensor networks. Also, they present an overview of key issues of WSNs and review literature of some aspects by classifying the problem into three groups. They review the major development of internal platform and its underlying operating system, communication protocol stack, and network services, provisioning, and deployment. In addition, these authors provide a discussion on the new challenges in the field of sensor networks.

WSNs target applications need a number of requirements which include range, antenna type, target technology, components, memory, storage, power, lifetime, security, computational capability, communication technology, size and programming interface. Interested readers may refer to the survey proposed in [4], in order to have information about some commercial and research sensors prototypes based on the above parameters. A comparison of presented sensors technologies against their addressable range, computational capability and storage capacity is provided in [4].

Researchers have provided a number of various contributions in order to achieve a large scale deployment and widespread management of WSNs in real applied domains. A recent applications of WSNs are highlighted in [69]. Authors propose the most recent technologies and testbeds for sensor networks which introduce appreciable synergies with others technologies. Based on that, they identify several challenges that need further investigations. In the same order, a selection of best testbeds applications of WSNs that have been developed for sensor evaluation is reviewed in [70].

Regarding consumption, the key challenges in WSNs is to maximize network lifetime while minimizing energy consumption. Techniques that allow the minimisation of energy wastage and the increasing of network lifetime by balancing the energy level of all nodes in network is reviewed in [62]. In the same order of ideas, the authors of [63] survey the main techniques used for energy conservation in sensor networks. They focused on duty cycling schemes that represent the most compatible technique for energy saving, data-driven approaches that can be used to improve the energy efficiency and on review of some communication protocols.

Limitation of resources in sensor nodes, security mechanism are difficult to implement in wireless sensor network. Because of that, it is necessary to neutralize attacks on all layers of sensor network protocol. More secured protocols are proposed [25, 27, 67, 68], but most of them introduce a lot of network overhead. An investigation of overhead due to the implementation of some common security mechanisms is given in [64]. Hence, providing a trusted WSNs remains a big challenge. A survey of some sensor networks security protocols is proposed by authors in [65, 66].

Except [61], which proposes an overview from application to physical layer, all of above presented survey are focused on a given aspect of sensor networks. Our survey is different from most proposed surveys in the literature in that, we discuss on all aspects and outlook of WSNs. We provide a comprehensive tool to get an overview and promising directions in sensor networks such as bio-inspired solutions which provide optimal methods for good management in sensor networks. To achieve that, we focused on most recent work in the literature.

## 3. SENSORS NETWORKS

A wireless sensor is a veritable embedded system with a wireless communication function, and that is capable to:

- Collect physical quantities such as heat, humidity, temperature, vibration, radiation, sound, light, movement, etc.
- Convert them into digital values which are sent as sensed data to a remote processing station or base station.

In general, there are two types of sensors: general's sensors and gateways sensors. A generic sensor has a role of collecting measures from the deployment area while the gateway sensor has more capacity in terms of computing resources, storage and transmission. Gateways sensors are generally used in a particular type of WSN architecture. A general architecture of WSN is given in figure 1.

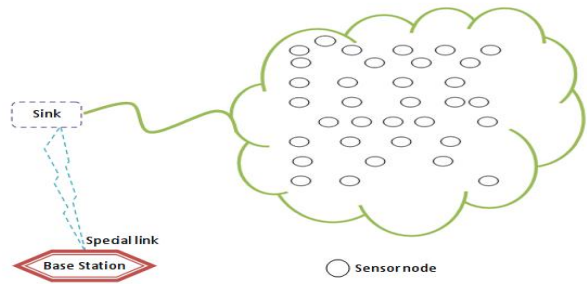

Figure 1. General architecture of WSN.

### 3.1. Wireless Sensor Networks

A WSN meanwhile can be defined as an adhoc network especially consisting of a number of wireless sensors that are deployed on a given area, to ensure an accurate task, either for monitoring or for tracking or for both. In [11], authors describe some aims of WSN design as following:

- self-organization, auto-recovery of faults, autonomous detection and correction of intrusions;
- scalability, adaptability, reliability and cooperative effort of sensor nodes;
- low power consumption, low node cost;
- routing, fault tolerance, QoS, security;
- survival to a change of topology in case of arrival and departure of node;
- survival to resources constraints.

Each sensor has an operating system. TinyOS is the specifically operating system designed for sensors and is thereby the most used [12, 13]. It's an event-driven operating system which provides a framework for programming embedded systems. The middle-ware proposed by TinyOS supports synchronization, routing, data aggregation, localization, radio communication, task scheduling, I/O processing, etc. Moreover, there are other sensors operating systems more or less popular: SOS cormos, EYES, PEEROS, MantisOS, Contiki, Kos, Senos, Nano-RK, LiteOS [14].

Particular type of ad hoc networks, WSNs have some differences with common ad hoc networks. Indeed, the number of nodes in a WSNs is more large (one to several thousand) and nodes are generally static and cooperate together to move gathered data towards the base station. In classical ad hoc networks, there are fewer nodes, but there's higher mobility. In terms of communication, WSNs have broadcasting mechanism throughout the network while classical ad hoc networks make the point to point communication. Also, in terms of energy consumption, it is lower in WSNs [15].

### 3.2. Types of WSNs

Depending on the deployment environment on earth, underground or underwater, there are several types of wireless sensor network.

- **Terrestrial**: In this type of sensor networks, hundreds to thousands of sensors deployed randomly or pre-deployed on a given area. This type of WSNs is mainly used in the field of environmental monitoring and presents a challenge to the sustainability of the network in terms of management of energy [16].

- **Underground**: These very special sensor nodes are known for their high cost and the required logistics for maintenance and pre-planned deployment. Sensors are installed in the soil for agriculture or in the walls of a mine to monitor conditions in the soil. However, in this type of network, there is land node which has role of relaying sensed information by the underground nodes to the base station [17].

- **Underwater**: This type of WSNs are still a great research challenge because of fact that environment in which the nodes are deployed is hostile and usually used for exploration. It's only possible to deploy a few nodes, these nodes are more expensive than terrestrial sensors, wireless communication is acoustic, the bandwidth is limited, the loss of signal is recurrent, propagation delays and synchronization problems are high [4, 19].

- **Multimedia**: This type of WSNs allows monitoring of a tracker in real-time events such as images, videos and sound. These sensors are equipped with cameras and microphones. Importance is given for: good bandwidth which implies a high energy consumption; processing and data compression; good QoS. Advance planning is necessary for the deployment of these sensors [20].

- **Mobile**: In this most recent type of WSNs, nodes are capable of repositioning and autonomously reorganize the network. After initial deployment, nodes disperse to collect information. There's also a hybrid network that consists of the combination of mobile sensors and fixed sensors [21].

## 4. ARCHITECTURE, OFFERED SERVICES AND FAULT TOLERANCE

Architecture of sensor network depends of expected service and the implemented application.

### 4.1. Architectures

Several sensor models are offered according to the underlying application. In general, a sensor node has: a sensor unit whose role is to capture a physical quantity and transform into a digital value; a data processing and storage unit; a wireless transmission module; a unit of management and control of energy. Also, depending on the application, some modules are added: GPS, solar cell, etc. Figure 2 present general node architecture.

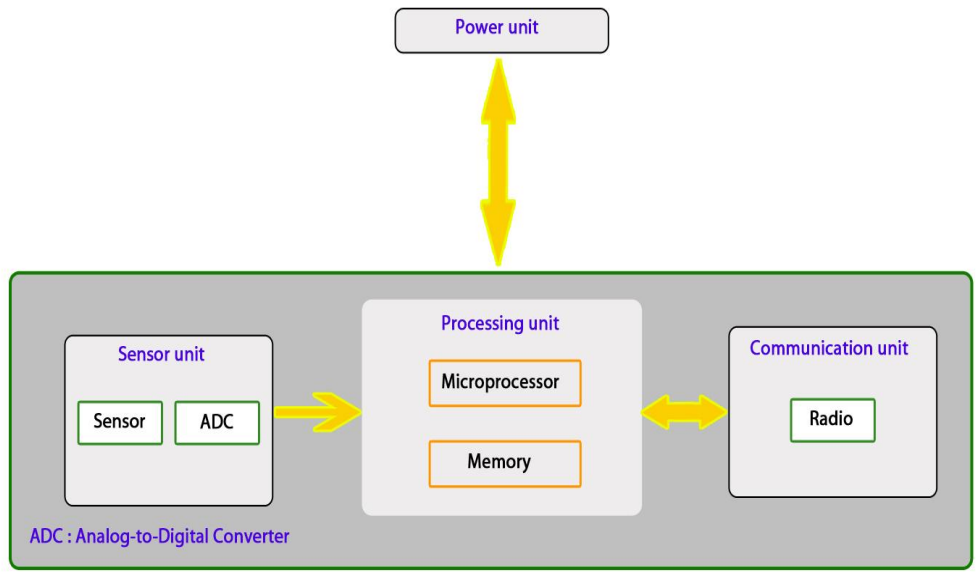

Figure 2. Sensor architecture.

There are two main types of network architectures for WSNs: (1) flat architecture, (2) hierarchical architecture.

1. In **flat architecture**, except for sink node, the other nodes are identical, they have the same capacity in terms of energy and computing, Also, they have the same role in sensing task. Node can directly communicate with a sink in single hop manner or communication with sink can be in multi hops manner.  Simplicity presented by this architecture enables low communication latency. Furthermore, when the network becomes denser, the scaling problem arises mainly as regards routing. Figure 3 present WSNs flat architectures.

2. A **hierarchical architecture** to deploy a large number of sensors. The network is divided into several groups or clusters which are the organizational unit of the network. Depending of cases, a more expensive cluster node type and more powerful than other nodes or a normal node in the cluster is designated as group leader called cluster head that is responsible of coordination of the sensors under its responsibility and act as a gateway to another cluster. The cluster head is responsible for the aggregation and/or compression of all the collected data in order to route it to the sink [22]. This allows the reduction of the transmission data within the network. However, there may be more latency in communications due to the density of the network and higher energy consumption for the cluster heads. Figure 4 highlight the clustered architectures for multi hops and single hop clusters.

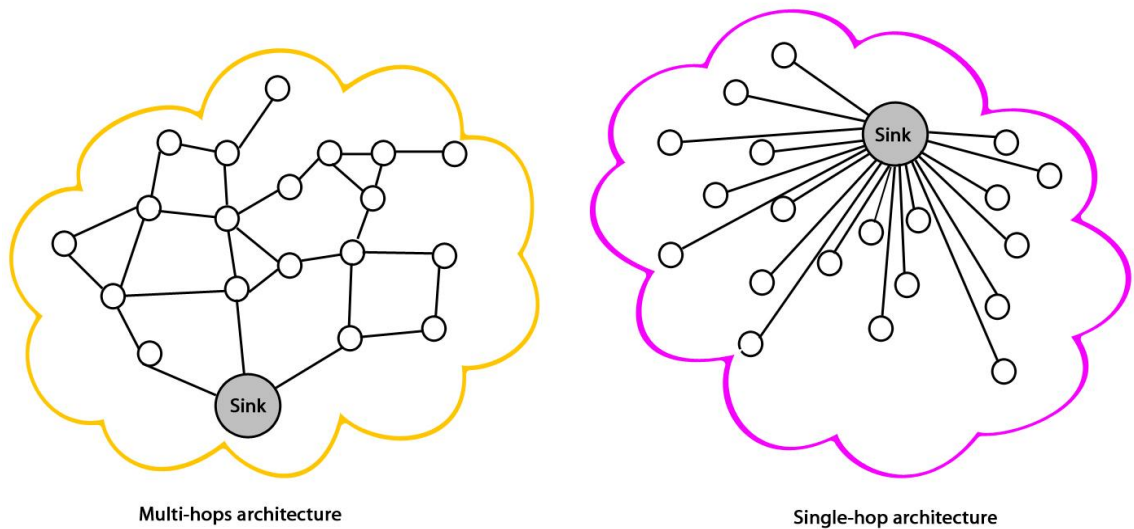

Figure 3. WSNs flat architectures.

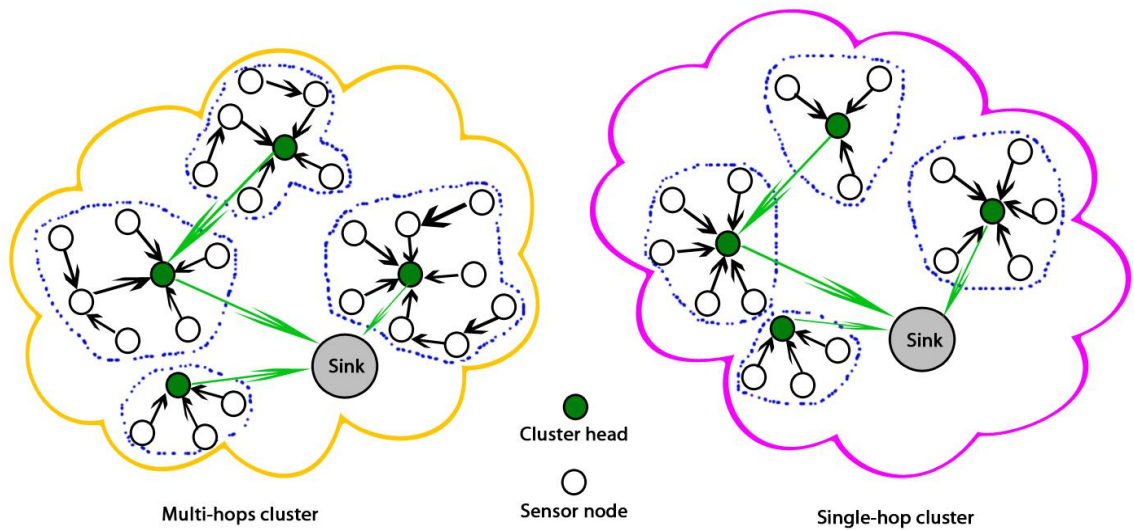

Figure 4. WSNs hierarchical architectures.

## 4.2. Offered services

Offered services in WSNs are organized into provisioning, control nodes management. Coverage and localization are the provisioning services. Control and nodes management enable WSN middle-ware to deliver services such as security, synchronization, compression and data aggregation.

1. **Coverage**: It involves the placement of nodes in deployment area to ensure coverage of the entire area by the network [28]. Coverage depends of application, the number of nodes and localization of these nodes.

2. **Localization**: It consists of finding position of a node in the network. In sensor network, three methods of localization are usually used [27] : - GPS (Global Positioning System) is the easiest, but cost of energy consumption is high and can't work on dense environment as forest; - anchor nodes approach which consist of node called anchor whose knows its position and help its neighboring nodes to evaluate their own; - close localization which consist of a node that using the neighboring nodes to determine their positions and then become an anchor node to other nodes.

3. **Synchronization**: Allow the realization of an activity or process at the same time as other nodes is an important service in WSNs. Synchronization is very important especially for routing and energy saving [26].

4. **Data compression**: It reduces the cost of communications and increase the reliability of data transfer. Compression involves the strict reduction of the amount of data to be routed towards the base station. Data compression requires decompression, which can reproduce compressed data in their original format. Decompression is achieved at the base station [24].

5. **Data aggregation**: It can greatly help to conserve the scarce energy resources by eliminating redundant data, thus achieving a longer network lifetime [23]. In other words, data aggregation is usually achieved by cluster head and consists of collecting data from all cluster members, applying an aggregation function to all collected data and transmits just one value as measured data.

6. **Data and network security**: WSNs are vulnerable to attack that consist of compromising a node, alter the integrity of the data, listening the network to retrieve messages or inject false messages in order to create a wasted of resources. Despite the difficult task of properly securing a WSN because of the significant limitation of resources, security protocols exist and ensure the security service nodes and network [22, 25].

**4.3. Fault tolerance in WSN**

Constraints in terms of resources capacities and the hostility of deployment environments expose WSNs to multiple to failures. Interactions with the environment or the deployment area may be the cause of network failures or any part of it. For a good scheduling operation, the network must be fault tolerant because sensors can have manufacturing error, lack of energy or victim of a security attack.

The aim of fault tolerance in WSNs is for example to achieve the network scheduling even if there's a node failure. In other words, fault tolerance in sensor network can be seen as ability to maintain network running without interruption in presence of failure of a sensor node. Fault tolerance is generally implemented in routing and transport protocols. It consists in the realization of the following two steps: detection of failure and limitation of effect; recovery and treatment of failure [38].

In particular, large-scale sensors networks are more exposed to failures because of inhospitable nature of their deployment environment. So restoring network after a damage is needed. According to this, authors in [60] propose a strategy for achieving fault tolerant network by establishing a bi-connected inter-cluster topology. The simulation of their approach shows a result in term of network connectivity.

In [39], a study of fault tolerant relay node placement problems is achieved and also discussion about complexity of proposed modified algorithms is done. In [40], authors present an artificial neural network model for fault tolerant WSN. A fault tolerant QoS clustering approach for WSNs is proposed in [41]. The use of dual cluster head mechanism guarantee the QoS, but some introduce security problem.

## 5. APPLICATIONS

The rapid evolution of sensor technology has led to design very small and smart sensors, which are used for various applications. Selection of a given type of sensor depends of desired application [29]. In fact, each application of WSN has a set of requested requirements as coverage, location, security, lifetime, etc. Classification of WSNs application can be mainly done into two categories: monitoring and tracking. In figure 5 we summarize this classification.

### 5.1. Monitoring

Monitoring is used to analyse, supervise and carefully control operations of a system or a process in real-time. Sensor network-based monitoring applications are various. Below some of them are briefly presented:

- **Environment**: monitoring of water quality, weather, pressure, temperature, seismic phenomena, vibration, monitoring of forest fires.
- **Agriculture**: irrigation management, humidity monitoring.
- **Ecology**: monitoring of animals in their natural environments.
- **Industry**: supply chain, inventory monitoring, industrial processes, machinery, productivity.
- **Smart house:** monitoring any addressable device in the house.
- **Urban**: transport and circulation systems, self-identification, parking management.
- **Health care**: organs monitoring, wellness, surgical operation.
- **Military**: intrusion detection.

### 5.2. Tracking

Tracking in WSN is generally used to follow an event, a person, animal or even an object. Existing applications in the tracking can be found in various fields.

- **Industry:** traffic monitoring, fault detection.
- **Ecology:** tracking the migration of animals in various areas.
- **Public health:** monitoring of doctors and patients in a hospital.

- **Military:** a WSN can be deployed on a battlefield or enemy zone to track, monitor and locate enemy troop movements.

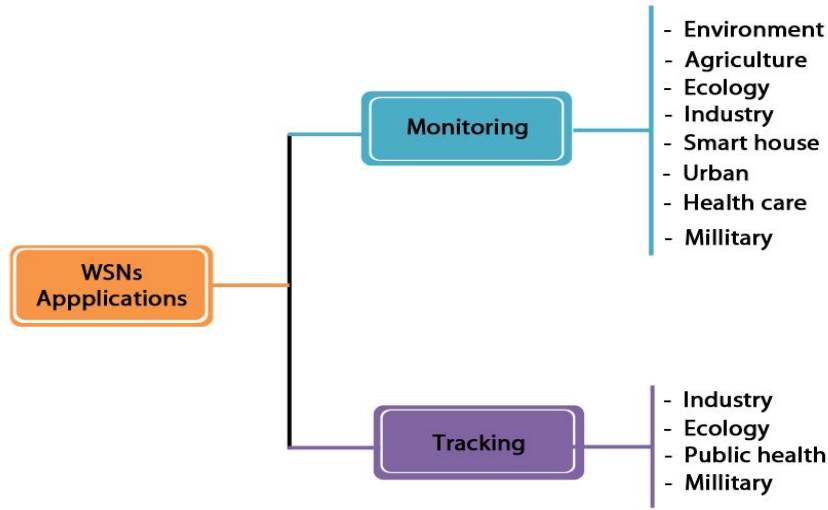

Figure 5. Classification of WSNs Applications.

### 5.3. Some practical applications

Precision agriculture is based on detailed processes of crop conditions such as the degree of fertilization, pesticide use or even crop protection against insects. A case of using WSN to achieve these advanced agricultural techniques was done in the south of Italy to produce tomatoes in a greenhouse. In fact, a sensor network is deployed for reducing pesticide usage in order to preserve environment and maximize the quality of tomatoes. The first application focused on measuring micro-climate of tomatoes crop to deliver detailed information for a novel decision system that help farmers to improve the quality of their production [30]. To deploy the WSN, Sensicast system is used and for management of network, the SensiNet platform was used [31].

In [32], authors present a case of using sensors in supply-chain for tracing transportation of perishable food. Indeed, to avoid the loss due to rotting in transportation, temperature sensors are deployed in trucks and other radio mechanisms called RFIDS are used for sending data and location to a remote site. In [33], authors has done an investigation of potential of sensor-based issuing policies such as FIFO, LIFO, SIRO, HQFO, LQFO, etc. on product quality in the perishables supply chain. In the same idea, authors in [34] have developed a real-time ZigBee based WSN for monitoring to the perishable food supply chain management. They also present their system architecture, hardware design and software implementation. The result shows a good network lifetime and high success rate in data transmission.

Sensors devices are also more used in medical monitoring system in order to improve health care of patients. There is a need of sensing applications in order to improve efficiency and quality of care in hospital environments. In [35], authors has deployed a WSN to monitor heart rate and blood oxygen levels at emergency room of a hospital and they use these data to known about the performance of hospital. Despite multiple frequency channels in hospital environment, their application achieve a good routing and high data reception rate.

Because climate change is a real problem for our planet, environmental monitoring become is an important goal. WSN is a domain which can provide solutions for real-time monitoring of environmental parameters as temperature or pressure in urban, rural, mountain or maritime zone. A case of study micro-climate phenomenon on top of a rock glacier in Switzerland by using WSN is described in [59]. Because of long-term experience nature of their project, authors have deployed a number of sensors with sufficient energy resources by adding secondary battery and solar panel. Environmental quantities measured by their WSN are: air humidity and temperature; precipitation; soil moisture; solar radiation; surface temperature; water content; wind direction and speed.

An application of sensor technology for environmental health monitoring in urban environments has been studied by authors in [36] for specific case in Nigeria country. Because of much environmental pollution in region of Niger Delta, authors have developed bio-monitoring approaches to study impacts on certain organisms. In the same idea, a study of monitoring water quality in urban reservoirs is presented in [37]. Authors propose an optimal sensor placement scheme to measure the wind distribution over a large urban reservoir with a limited number of wind sensors.

## 6. COMMUNICATION PROTOCOLS

The communication protocols in WSNs are different from traditional communication protocols because of the strong limitation sensor resources [18]. These protocols are based on five layers: application, transport, network, link and physical. Depending on the gathered value, software is able to exploit data from application layer in order to compute and interpret the collected data. The transport layer ensures the reliability and quality of data between source and destination. Network layer in collaboration with transport layer, has a role of routing data across the network. The link layer allows detection and error correction, and contributes to the reduction of the collision of messages in the network. The physical layer provides an interface to send/receive byte stream to/from the communication channel.

Network protocols depend largely on the transport protocol. Implementation of transport protocol must be general and independent of the application. Transport protocol manages congestion in the network, transmission reliability and ensures energy conservation. In the field of WSNs, routing data from a source to a destination is a task that requires a build of fault tolerant, secure and fairness protocol [29]. Routing protocol in WSNs can be categorized in four groups: Data-centric protocol; Hierarchical protocol; Location-based protocol and Bio-inspired protocol.

**Data-centric protocols:** The main concept implemented by these categories of network protocol is to control and eliminate redundant data in the network.

**Hierarchical protocols**: These kinds of protocol are generally designed for large scale WSNs. Nodes are organized into clusters. Each cluster head is responsible to aggregate data for transmission to the base station. This is done in order to reduce the energy consumption of sensor nodes.

**Location-based protocols:** These protocol categories use the position information to send the data only to the desired destination.

**Bio-inspired protocols:** These types of routing protocol are more recent. It consists of using analogies between computing methods and biological behaviors of swarms in which collective intelligence can emerge. These swarms are colonies of social insects, bird flocking and fish schooling, firefly, etc...

In table 1, we present a comparison of some routing protocols.

| | Routing | Scalability | Synchronization | Coverage | Data aggregation | Security | Overhead | Energy-consumption | Maintenance |
|---|---|---|---|---|---|---|---|---|---|
| HEERP [42] | Hierarchical | yes | yes | - | yes | - | less | less | yes |
| EADC [43] | Hierarchical | yes | yes | yes | yes | - | less | less | - |
| U-LEACH [44] | Hierarchical | yes | yes | - | yes | - | less | less | |
| ALS [45] | Location-based | yes | no | - | no | no | medium | - | - |
| [46] | Location-based | yes | no | - | no | no | medium | - | - |
| MSDD [47] | Data-centric | no | yes | yes | no | no | medium | less | yes |
| [48] | Data-centric | no | no | no | no | no | high | high | - |
| [49] | Data-centric | no | no | no | no | no | medium | less | - |
| [50] | Bio-inspired | yes | no | yes | yes | - | less | less | yes |
| [51] | Bio-inspired | yes | no | yes | no | - | less | less | yes |

## 7. SECURITY IN SENSOR NETWORKS

In order to design a WSN application, it is supposed that all sensor nodes are each other worthy of trust. However, sensor nodes are generally deployed on uncontrolled and inhospitable environments. This situation exposes the sensors to different kinds of attacks that can totally damage network operations. Indeed, these attacks mainly exploit the uncertainty of the communication channel and the random deployment of sensors on an uncontrolled area. Thus, ensure the safety of this type of network is a difficult task, especially because nodes have limited hardware capabilities [71].

Security in WSNs can be classified into two broad categories: operational security and information security. The security of WSNs can be classified into two broad categories: QoS and security. The first category aim to ensure the continuity of operations in the entire network, even if, there is a faulty node or if a node was attacked. For the second, the objective is to ensure data confidentiality, integrity, authentication, availability and freshness. In fact, an attacker can compromise a sensor node by altering the integrity of the data, injecting fake data on the network or eavesdropping. These attacks are commonly partitioned into physical and logical vulnerabilities.

Physical vulnerability is a kind of attack in which an adversary alter a part of sensor, such as changing its programming code or replace a given sensor by a compromised node. Logical vulnerabilities lie in the programs and protocols. Furthermore, some attacks intended to affect the integrity of messages that pass through the network, while others are designed to reduce the availability of the network or its components. These attacks are in two kinds: passive and active.

In passive attack, the goals of adversary is to collect information in the network without being discovered [72]. This is possible because of technology of wireless communication channel. In fact, transmissions are broadcast by radio waves, no network access control is possible. The most known attack based on that is called eavesdropping. Therefore, it's very easy to intercept exchanged data and analyse the traffic if there is no planned privacy service.

Active attacks are more harmful than passive attacks, for network operations and lifetime. When an attacker successfully compromises the network, he can modify messages, introduce unneeded traffic in order to exhaust node energy. In this range are Wormhole, Sybil and Sinkhole attack, that are known as routing attack because they act on network layer [72]. Details of attacks per layer can be found on [72, 73]. In the same order of harm, Denial of Service (DoS) attacks which consist of sending an unlimited number of messages in order to exhaust resources, are implemented in different layer of protocol stack.

Due to resources limitation, it's therefore necessary to design new robust algorithms to carry out routing operations even in the presence of malicious nodes. In addition, securing data aggregation operations and node localization schemes remains a challenge. Even so, various solutions are proposed for designing secure sensor networks [23, 68, 75]. Also, bio-inspired techniques promise an outlook for secure sensor networks. For example, a bio-inspired cryptographic algorithm for WSNs based on genetic programming is proposed in [74].

## 8. OPEN RESEARCH ISSUES

In general, the challenges are the designing of WSNs by taking in account of limited energy capacity, resources constraints, random and large deployment, dynamic and not controlled environment. In this section, we propose a small discussion about the opened research issues in terms of services, applications, fault tolerance and communication protocols. Also, discuss on bio-inspired solutions which promise a good perspective in fields of WSNs.

Current localization algorithms still have a considerable cost in terms of energy consumption [27]. It's therefore, necessary to design energy efficient localization algorithms and techniques. Promising nature-inspired approaches for nodes localization are proposed in [52, 53]. The proposed algorithms show a little gain in term of energy consumption, but improvement remain possible.

In another hand, WSN services such as coverage, synchronization, collision control, data aggregation and compression, are still a challenge which need particular attention for better improvements. Detection and response to an attack without interrupting the operation of the network is a challenge for secure networks. In particular, for security issues and because data is the final need by applications, secure data become much important for on top applications. In fact, a compromised node may send arbitrary data and delude the sink node. In [2], a robust adaptive approach based on hierarchical monitoring providing trust data aggregation is proposed to secure data received by the base station. Also, secure all WSNs services remain a big challenge because of sensor resource constraints.

Otherwise, the rapid development of wireless sensor technology is mainly due to the different needs of applications. Also, each sensors networks application is specific to a given domain. Therefore, there is a need of development of application solutions by taking in account of specificities of concerned domain. These specificities must meet the constraints of sensor technology and therefore this leads to research questions.

Enable reliable communication while managing congestion in the network is the main challenge of the transport layer on sensor protocol stack. In terms of network protocols, several are offered depending on the considered sensor network architecture. However, fairness, reliability, congestion control and overhead are still problems to solve in network protocols [54]. In order to optimize network protocols, introduction of cross-layer approach are proposed for WSNs. In fact, interaction between all layers of the protocol stack may optimize sensors performances and so cross-layer optimization becomes an important research direction for sensor network. A case of optimizing communication protocol by using cross-layer optimization combined with a Markov chain model is proposed in [55].

Moreover, bio-inspired solutions are progressively applied in WSNs [50, 51]. Swarms intelligence, artificial immune system, genetic algorithm and other nature-inspired phenomenon in which self-organization and collective intelligence can emerge, are increasingly used in the fields of sensor networks in order to optimize the whole management of WSNs [56, 57, 58]. We believe that more nature-inspired applications and models will be proposed in the near future and this will provide a good progress in WSN.

## 9. CONCLUSION

WSNs offer great opportunities for diverse applications. This technology and existing applications are constantly growing. Also the research in this field is very exciting because of a variety of offered services by sensors networks. In this paper, we present a review of WSNs by highlighting some aspects of this particular type of ad hoc networks, and open research directions in this field. Nevertheless, several problems still require improvements, despite the number of contributions. In particular, security, localization, energy efficient intelligent routing, etc. remain open. Also, we made a small discussion about the promise of biological inspired solutions which provide a good management of sensor network. We believe that bio-inspired solutions constitute an interesting area for further research for optimizing WSNs services.